\documentclass[12pt]{article}

\usepackage{a4wide}
\usepackage[pdftex,usenames,dvipsnames]{color}
\usepackage{graphics}
\usepackage{amsfonts}

\newcommand{\nit}{\noindent}

\newcommand{\np}{\newpage}
\newcommand{\dsp}{\displaystyle}
\newcommand{\vs}[1]{\vspace{#1 ex}}
\newcommand{\hs}[1]{\hspace{#1 em}}
\newcommand{\bfr}{\begin{flushright}}
\newcommand{\efr}{\end{flushright}}
\newcommand{\bc}{\begin{center}}
\newcommand{\ec}{\end{center}}
\newcommand{\ben}{\begin{enumerate}}
\newcommand{\een}{\end{enumerate}}

\newcommand{\be}{\begin{equation}}
\newcommand{\ee}{\end{equation}}
\newcommand{\ba}{\begin{array}}
\newcommand{\ea}{\end{array}}

\newcommand{\ct}{\cite}

\newcommand{\ag}{\alpha}
\newcommand{\bg}{\beta}
\newcommand{\gam}{\gamma}
\newcommand{\del}{\delta}

\newcommand{\ve}{\varepsilon}
\newcommand{\zg}{\zeta}
\newcommand{\thg}{\theta}
\newcommand{\kg}{\kappa}
\newcommand{\lb}{\lambda}
\newcommand{\sg}{\sigma}
\newcommand{\rg}{\rho}

\newcommand{\vf}{\varphi}
\newcommand{\og}{\omega}
\newcommand{\Gam}{\Gamma}
\newcommand{\Del}{\Delta}

\newcommand{\Sg}{\Sigma}

\newcommand{\bfJ}{\bold {J}}

\newcommand{\lh}{\left(}
\newcommand{\rh}{\right)}
\newcommand{\ld}{\left.}
\newcommand{\rd}{\right.}

\newcommand{\nb}{\nabla}

\newcommand{\ctg}{\mbox{\,cotan\,}}

\newcommand{\der}{\partial}

\begin{document}

\pagestyle{empty}
\bfr
Nikhef-2015-041
\efr

\bc
{\Large {\bf Spinning bodies in curved space-time}}\\
\vs{7}

{\large G.\ d'Ambrosi$^{a \ast}$, S.\ Satish Kumar$^{b \dagger}$}\\
\vs{2}

{\large J.\ van de Vis$^{a,b \ddagger}$, J.W.\ van Holten$^{a,b \ast\ast}$}
\vs{5}

\today
\ec
\vs{5}

\nit
{\footnotesize {\bf Abstract} \\
We study the motion of neutral and charged spinning bodies in curved space-time in the test-particle limit. We construct 
equations of motion using a closed covariant Poisson-Dirac bracket formulation which allows for different choices of the 
hamiltonian. We derive conditions for the existence of constants of motion and apply the formalism to the case of 
spherically symmetric space-times. We show that the periastron of a spinning body in a stable orbit in a Schwarzschild 
or Reissner-Nordstr{\o}m background not only precesses, but also varies radially. By analysing the stability conditions 
for circular motion we find the innermost stable circular orbit (ISCO) as a function of spin. It turns out that there is an 
absolute lower limit on the ISCOs for increasing prograde spin. Finally we establish that the equations of motion can 
also be derived from the Einstein equations using an appropriate energy-momentum tensor for spinning particles. }

\vfill

\footnoterule 
\nit
{\footnotesize $^a$ Nikhef, Science Park 105, Amsterdam NL }\\
{\footnotesize $^b$ Lorentz Institute, Leiden University, Niels Bohrweg 2, Leiden NL}\\
{\footnotesize \begin{tabular}{ll} $^\ast$ {\it gdambros@nikhef.nl} & $^\dagger$ {\it satish@lorentz.leidenuniv.nl} \\
     $^{\ddagger}$ {\it jorindev@nikhef.nl} & $^{\ast\ast}$ {\it v.holten@nikhef.nl}  \end{tabular} }

\np
\pagestyle{plain}
\pagenumbering{arabic}

\section{Introduction \label{ps1}}

Relativistic dynamics is becoming increasingly relevant in astrophysics and cosmology, for example when it 
comes to understanding compact stars, black holes and gravitational waves. Since long it was realized that 
spin, i.e.\ the internal angular momentum of compact objects, can have important dynamical effects. As a 
consequence there is an extensive literature on the subject 
\ct{deSitter:1916zza,Thomas:1926dy,Frenkel:1926zz,Mathisson:1937zz,Papapetrou:1951pa,Fock:1954,
Dixon:1970zza,Wald:1972sz,Hanson:1974qy,Semerak:1999qc}. 

In a recent letter \ct{d'Ambrosi:2015gsa} three of us presented a non-standard covariant description of 
spinning bodies in the test-particle limit. In contrast to the most-widely used approach of test-particle 
dynamics \ct{Schaefer:2004qh,Semerak:2007,Plyatsko:2011gf,Steinhoff:2010zz,Costa:2012cy} it is 
formulated in terms of a covariant kinetic momentum rather than a canonical momentum which is not 
proportional to the four-velocity of the body. The price to pay is a world-line which does not always 
coincide with that of a center of mass; rather it follows the spin, with the result that there is a mass 
dipole describing the displacement between the two in the presence of curvature. There are several 
advantages to this formulation: it does not require constraints like the Pirani or Tulczyjew conditions, 
it can be used with a variety of hamiltonians and it makes the analysis of the motion of spinning 
test-particles in curved space-time more tractable; in some cases of practical interest exact results 
are obtained.

In this paper we continue to develop this formalism of spin-dynamics in the test-particle limit. It is 
organised as follows. In section \ref{ps2} we provide a brief summary of the hamiltonian phase-space 
approach of reference \ct{d'Ambrosi:2015gsa}, extending it to include also electric charge and 
electromagnetic fields. We proceed with the role of Killing vectors in constructing constants of motion 
in section \ref{ps3}. In section \ref{ps4} the formalism is applied to motion associated with a minimal 
kinetic hamiltonian in static and spherically symmetric space-times of Schwarzschild and 
Reissner-Nordstr{\o}m type. We analyse circular orbits in section \ref{ps5} and find non-circular orbits 
by a perturbative construction in section \ref{ps6}. We also determine the effects of spin on the periastron 
and show that in addition to precessing the periastron also shifts radially. In section \ref{4pisco} the stability 
of circular orbits is analysed in detail to determine the radius of the innermost stable circular orbit (ISCO) 
as a function of spin. The stability conditions are found to impose an upper limit on the spin associated 
with the ISCO. Section \ref{ps7} describes how to include Stern-Gerlach forces 
\ct{Khriplovich:2008ni,Khriplovich:1998ev,Khriplovich:1997ni,Khriplovich:1996dv,Khriplovich:1989ed}, 
both of electromagnetic and of gravitational origin, using a class of non-minimal hamiltonians. We show 
that they allow for an extension of all the constants of motion associated with the minimal hamiltonian 
and determine also the circular orbits for this case. In section \ref{ps8} it is shown that the equations of 
motion we use can be derived in a different way from consistency of the Einstein equations with an 
appropriate energy-momentum tensor. Finally in section \ref{ps9} we conclude with a discussion and 
summary. Some mathematical details have been collected in the appendices.

\section{Covariant Hamilton formalism for spinning bodies \label{ps2}}

Test-particles are gravitational objects characterised by a finite number of degrees of freedom such as 
mass, charge and spin, of which the back reaction on space-time geometry can be considered negligible. 
Thus the phase space is finite-dimensional; it is spanned by the position $x^{\mu}$, momentum $\pi_{\mu}$ 
and anti-symmetric spin-tensor $\Sg^{\mu\nu}$, with mass $m$ and charge $q$ constant parameters 
characterizing the strength of interactions. Whilst the four-velocity $u^{\mu} = \dot{x}^{\mu}$ is a time-like unit 
vector, the spin-tensor can actually be decomposed in a space-like vector $Z^{\mu}$, the mass dipole vector, 
and a space-like axial vector $S_{\mu}$, the spin proper:
\be
\Sg^{\mu\nu} = - \frac{1}{\sqrt{-g}}\, \ve^{\mu\nu\kg\lb} u_{\kg} S_{\lb} + u^{\mu} Z^{\nu} - u^{\nu} Z^{\mu},
\label{2.0}
\ee
where
\be
S_{\mu} = \frac{1}{2}\, \sqrt{-g}\, \ve_{\mu\nu\kg\lb} u^{\nu} \Sg^{\kg\lb}, \hs{1}
Z^{\mu} = \Sg^{\mu\nu} u_{\nu},
\label{2.1}
\ee
such that $S \cdot u = Z \cdot u = 0$. It is interesting to note that in addition one can define a third 
space-like vector
\be
W^{\mu} = - \frac{1}{\sqrt{-g}}\, \ve^{\mu\nu\kg\lb} u_{\nu} S_{\kg} Z_{\lb} 
 = \lh \Sg^{\mu\nu} - u^{\mu} Z^{\nu} \rh Z_{\nu},
\label{2.2}
\ee
orthogonal to the other ones:
\be
W \cdot u = W \cdot S = W \cdot Z = 0.
\label{2.3}
\ee
Together $(u, S, Z, W)$ form a set of independent vectors, one time-like and three space-like, 
which can be used to define a frame of basis vectors carried along the particle world-line.  

We build our formulation of the dynamics on a set of covariant Poisson-Dirac brackets which are closed 
in the sense of Jacobi identies and independent of the specific hamiltonian:
\be
\ba{l}
\left\{ x^{\mu}, \pi_{\nu} \right\} = \del^{\mu}_{\nu}, \hs{2}
\left\{ \pi_{\mu}, \pi_{\nu} \right\} = \frac{1}{2}\, \Sg^{\kg\lb} R_{\kg\lb\mu\nu} + q F_{\mu\nu}, \\
 \\
\left\{ \Sg^{\mu\nu}, \pi_{\lb} \right\} = \Gam_{\lb\kg}^{\;\;\;\mu}\, \Sg^{\nu\kg} - \Gam_{\lb\kg}^{\;\;\;\nu}\, \Sg^{\mu\kg}, \\
 \\
\left\{ \Sg^{\mu\nu}, \Sg^{\kg\lb} \right\} = g^{\mu\kg} \Sg^{\nu\lb} - g^{\mu\lb} \Sg^{\nu\kg} - g^{\nu\kg} \Sg^{\mu\lb}
 + g^{\nu\lb} \Sg^{\mu\kg},
\ea
\label{2.4}
\ee
all other brackets vanishing. The structure functions appearing in these brackets are the metric, connection
and curvature tensor of the space-time manifold, with the electromagnetic field strength $F_{\mu\nu}$ 
appearing as the structure function for the central charge $q$.

Next we have to specify a hamiltonian to generate the equations of motion via the brackets (\ref{2.4}).
The minimal choice is the kinetic hamiltonian
\be
H_0 = \frac{1}{2m}\, g^{\mu\nu}(x) \pi_{\mu} \pi_{\nu}. 
\label{2.5}
\ee
Other choices are possible, and we will discuss a specific example in section \ref{ps7}. 
The equations of motion derived from the hamiltonian $H_0$ are 
\be
\pi_{\mu} = g_{\mu\nu} m u^{\nu}, \hs{2} 
m D_{\tau} u^{\mu} = \frac{1}{2}\, \Sg^{\kg\lb} R_{\kg\lb\;\,\nu}^{\;\;\;\,\mu} u^{\nu} + q F^{\mu}_{\;\,\nu}\, u^{\nu},
\label{2.6}
\ee
for the world-line, and 
\be
D_{\tau} \Sg^{\mu\nu} = 0,
\label{2.7}
\ee 
for the spin-tensor; here $D_{\tau}$ is the pullback of the covariant derivative on the world-line. It follows that
the world-line is a curve on which the spin-tensor is covariantly constant. This does not hold for the vectors 
$S$ and $Z$ individually as they satisfy 
\be
\ba{l}
\dsp{ m D_{\tau} S_{\mu} = \frac{1}{2}\, \sqrt{-g}\, \ve_{\mu\nu\kg\lb} \Sg^{\kg\lb} 
  \lh \frac{1}{2}\, \Sg^{\ag\bg} R_{\ag\bg\;\,\sg}^{\;\;\;\;\nu} + q F^{\nu}_{\;\sg} \rh u^{\sg}, }\\
 \\
\dsp{ m D_{\tau} Z^{\mu} = \Sg^{\mu\nu} \lh \frac{1}{2}\, \Sg^{\kg\lb} R_{\kg\lb\nu\sg}+ q F_{\nu\sg} \rh u^{\sg}. }
\ea
\label{2.8}
\ee
Comparing with the conventional analysis of spinning particle dynamics 
\ct{Mathisson:1937zz,Papapetrou:1951pa,Dixon:1970zza} it is seen that the constraint on the mass dipole (e.g.\
the Pirani condition \ct{Pirani:1956tn}) has been replaced here by a proper equation of motion. As a result in our
formulation of classical relativistic mechanics the mass-dipole $Z$ cannot vanish permanently, except in flat 
field-free Minkowski space-time.\footnote{However one can have $Z^{\mu} = 0$ in pseudo-classical models in 
which the spin-tensor is represented in terms of Grassmann variables $\Sg^{\mu\nu} = i \psi^{\mu} \psi^{\nu}$, 
because of conservation of the supercharge $Q = \psi^{\mu} \pi_{\mu}$ 
\ct{Berezin:1976eg,Brink:1976uf,vanHolten:1992bu}.} However it is easily established that the two approaches 
can be made to agree to linear order in the spin tensor.

\section{Constants of motion \label{ps3}}

In the hamiltonian formalism constants of motion are found by requiring its bracket with the hamiltonian to vanish.
There are three generic constants of motion for any space-time geometry. First the hamiltonian itself, which defines 
the particle mass:
\be
H_0 = - \frac{m}{2} \hs{1} \Rightarrow \hs{1} g_{\mu\nu} u^{\mu} u^{\nu} = -1.
\label{3.1}
\ee
In addition there are two constants of motion for the spin: the total spin 
\be
I = \frac{1}{2}\, g_{\mu\kg} g_{\nu\lb} \Sg^{\mu\nu} \Sg^{\kg\lb} 
 = S \cdot S + Z \cdot Z,
\label{3.2}
\ee
and the pseudo-scalar product 
\be
D = \frac{1}{8}\, \sqrt{-g}\, \ve_{\mu\nu\kg\lb} \Sg^{\mu\nu} \Sg^{\kg\lb} = S \cdot Z.
\label{3.3}
\ee
Furthermore there are constants of motion depending on the symmetries of the geometry. They are 
constructed in terms of Killing vectors and tensors. In particular constants of motion $J$ of the form
\be
J = \gam + \ag^{\mu} \pi_{\mu} + \frac{1}{2}\, \bg_{\mu\nu} \Sg^{\mu\nu},
\label{3.4}
\ee
exist if 
\be
\nb_{\mu} \ag_{\nu} + \nb_{\nu} \ag_{\mu} = 0, \hs{1} \nb_{\lb} \bg_{\mu\nu} = R_{\mu\nu\lb\kg} \ag^{\kg}, 
\hs{1} \der_{\mu} \gam = q F_{\mu\nu} \ag^{\nu}.
\label{3.5}
\ee
Thus $\ag^{\mu}$ is a Killing vector and $\bg_{\mu\nu}$ its curl:
\be
\bg_{\mu\nu} = \frac{1}{2} \lh \nb_{\mu} \ag_{\nu} - \nb_{\nu} \ag_{\mu} \rh,
\label{3.6}
\ee
whilst a solution for $\gam$ can be found if the Lie-derivative of the vector potential with respect to $\ag$ 
vanishes:
\be
\ag^{\nu} \der_{\nu} A_{\mu} + \der_{\mu} \ag^{\nu} A_{\nu} = 0 \hs{1} \Rightarrow \hs{1} 
\gam = q A_{\mu} \ag^{\mu}.
\label{3.7}
\ee
This requirement in fact states that the electromagnetic and gravitational fields must both exhibit the same 
symmetries for an associated constant of motion to exist.

\section{Spherical symmetry \label{ps4}}

There are few static solutions of the Einstein equations possessing spherical symmetry. The most relevant 
ones are the Minkowski and Schwarzschild geometries. In addition the Reissner-Nordstr{\o}m geometry is 
a static and spherically symmetric solution of the Einstein-Maxwell equations. In these symmetric space-times
the orbital angular momentum combines with the spin to create a conserved total angular momentum. 
In this section we consider the exterior Reissner-Nordstr{\o}m geometry of an electrically charged massive
spherical body and its reduction to Schwarzschild space-time in the limit of vanishing charge. 

The standard form of the Schwarzschild-Reissner-Nordstr{\o}m metric is represented by the line element
\be
g_{\mu\nu} dx^{\mu} dx^{\nu} = - \lh 1 - \frac{2M}{r} + \frac{Q^2}{r^2} \rh dt^2 + 
 \frac{dr^2}{1 - \frac{2M}{r} + \frac{Q^2}{r^2}} + r^2 d\thg^2 + r^2 \sin^2 \thg\, d\vf^2.
\label{4.1}
\ee
Here $M$ is the mass and $Q$ is the charge of the source creating the space-time curvature as well as a 
static electric Coulomb field 
\be
A = A_{\mu} dx^{\mu} = - \frac{Q}{r}\, dt, \hs{2} \frac{1}{2}\, F = dA = \frac{Q}{r^2}\, dr \wedge dt.
\label{4.2}
\ee
The components of the connection and curvature tensor are collected in appendix \ref{aa}. In the line 
element (\ref{4.1}) $t$ and $\vf$ are cyclic co-ordinates implying simple constant Killing vectors for time 
translations and rotations around the $z$-axis. Two more Killing vectors exist generating rotations 
around the two other axes. As also the Coulomb potential is static and spherically symmetric we obtain 
four constants of motion: \\
- the particle energy 
\be
E = - \pi_t  + \frac{qQ}{r} - \lh \frac{M}{r^2} - \frac{Q^2}{r^3} \rh \Sg^{tr},
\label{4.3}
\ee
with $Q = 0$ in Schwarzschild space-time; \\
- and the 3-vector of total angular momentum
\be
\ba{lll}
J_1 & = & \dsp{ - \sin \vf\, \pi_{\thg} - \ctg \thg \cos \vf\, \pi_{\vf} }\\
 & & \\
 & & \dsp{ - r \sin \vf\, \Sg^{r\thg} - r \sin \thg \cos \thg \cos \vf\, \Sg^{r\vf} + r^2 \sin^2 \thg \cos \vf\, \Sg^{\thg\vf}, }\\
 & & \\
J_2 & = & \dsp{ \cos \vf\, \pi_{\thg} - \ctg \thg \sin \vf\, \pi_{\vf} }\\
 & & \\
 & & \dsp{ + r \cos \vf\, \Sg^{r\thg} - r \sin \thg \cos \thg \sin \vf\, \Sg^{r\vf} + r^2 \sin^2 \thg \sin \vf\, \Sg^{\thg\vf}, }\\
 & & \\
J_3 & = & \dsp{ \pi_{\vf} + r \sin^2 \thg\, \Sg^{r\vf} + r^2 \sin \thg \cos \thg\, \Sg^{\thg\vf}. }
\ea
\label{4.4}
\ee
The analysis of the equations of motion can be simplified considerably by using the spherical symmetry 
to fix the direction of the total angular momentum to be the $z$-axis
\be
\bfJ = (0, 0, J), \hs{1} J = m r^2 u^{\vf} + r \Sg^{r\vf},
\label{4.5}
\ee
resulting in
\be
\Sg^{r\thg} = - m r u^{\thg}, \hs{1} \Sg^{\thg\vf} = \frac{J}{r^2}\, \ctg \thg.
\label{4.6}
\ee
Clearly this choice does not fix the motion to be in the equatorial plane $\thg = \pi/2$, as the spin and 
orbital angular momentum are not necessarily aligned; indeed it is possible to have precession of 
both the spin and the orbital angular momentum, making the plane of the orbit precess \ct{Bini:2014poa} as well.

\section{Circular orbits \label{ps5}} 

In a Schwarzschild-Reissner-Nordstr{\o}m background planar motion requires alignment of the spin 
and orbital angular momentum. In this section we consider in particular the conditions for plane circular 
motion. The analysis is most conveniently done by taking the plane of motion to be the equatorial plane, 
which implies $\thg = \pi/2$ and $u^{\thg} = 0$ as well as $\dot{u}^{\thg} = 0$. As a result
\be
\Sg^{t\thg} = \Sg^{r\thg} = \Sg^{\thg\vf} = 0.
\label{5.1}
\ee
From the definition (\ref{3.3}) it follows that such orbits satisfy the conservation law
\be
D = S \cdot Z = 0.
\label{5.0}
\ee
For circular motion we additionally impose $r = R =$ constant and $u^r = \dot{u}^r = 0$. 

\nit
Using the equation of motion (\ref{2.6}) for the $r$-co-ordinate, the latter condition leads to
\be
\ba{l}
\dsp{ \lh 2MR^3 - 3M^2R^2 - 3 R^2Q^2 + 6MRQ^2 - 2Q^4 \rh \lh R^2 - 2MR + Q^2 \rh u^{t\,2} }\\
 \\
\dsp{ \hs{3} -\, \lh R^2 - 3MR + 2Q^2 \rh \lh MR - Q^2 \rh R^4 u^{\vf\,2} = }\\
 \\
\dsp{ \left[ \lh 2MR - 3Q^2 \rh \ve - \lh MR - 2Q^2 \rh \frac{qQ}{mR} \right] \lh R^2 - 2MR + Q^2 \rh R^2 u^t
 + \lh MR - Q^2 \rh^2 \eta R^2 u^{\vf}, }
\ea
\label{5.2}
\ee
where $\ve = E/m$ and $\eta = J/m$ are the particle energy and angular momentum per unit of mass. 
The hamiltonian constraint (\ref{3.1}) for circular orbits becomes
\be 
\lh 1 - \frac{2M}{R} + \frac{Q^2}{R^2} \rh u^{t\,2} = 1 + R^2 u^{\vf\,2}.
\label{5.3}
\ee
Hence (\ref{5.2}) and (\ref{5.3}) constitute two independent equations for $u^t$ and $u^{\vf}$ in terms of 
$(M,Q,R)$ and $(\ve,\eta)$; therefore $u^t$ and $u^{\vf}$ are constant on circular orbits: 
\be
\dot{u}^t = \dot{u}^{\vf} = 0.
\label{5.4}
\ee
As a result on circular orbits the equations of motion for $u^t$ and $u^{\vf}$ reduce to
\be
m g_{tt} \dot{u}^t = \Sg^{t\vf} R_{t\vf t\vf} u^{\vf} = 0, \hs{1}
m g_{\vf\vf} \dot{u}^{\vf} = \Sg^{t\vf} R_{t\vf t\vf} u^t = 0,
\label{5.5}
\ee
implying that $\Sg^{t\vf} = 0$. This in turn also requires $\dot{\Sg}^{t\vf} = 0$, or
\be
\lh MR - Q^2 \rh u^t \Sg^{r\vf} + \lh R^2 - 2MR + Q^2 \rh u^{\vf} \Sg^{tr} = 0.
\label{5.6}
\ee
Using the conservation laws for $E$ and $J$ this can be replaced by
\be
\ba{l}
\dsp{ \lh MR - Q^2 \rh^2 \eta u^t - \lh R^2 - 2MR +Q^2 \rh \lh \ve - \frac{qQ}{mR} \rh R^4 u^{\vf} }\\
 \\
\hs{4} + \lh R^2 -3MR + 2Q^2 \rh \lh R -M \rh R^3 u^t u^{\vf} = 0.
\ea
\label{5.7}
\ee
As this is yet another equation between $(u^t,u^{\vf})$ and the constants of motion, it fixes a relation
between the energy $E = m \ve$ and total angular momentum $J = m \eta$ for a given circle $r = R$. 
In fact, we can now eliminate $\eta$ between eqs.\ (\ref{5.2}) and (\ref{5.7}), using the constraint 
(\ref{5.3}), to get 
\be
\ba{l}
\lh R^3 - 3MR^2 + 3M^2R - M Q^2 \rh u^{t\,3} - \lh R^2 - 3MR + 2Q^2 \rh R u^t  \\
 \\
\dsp{ \hs{2} =\, - R^3 \lh \ve - \frac{qQ}{mR} \rh 
 + \left[ \lh  R^2 - 2Q^2 \rh \ve - \lh R^2 - MR - Q^2 \rh \frac{qQ}{mR} \right] R u^{t\,2}. }
\ea
\label{5.8}
\ee
For the case of Schwarzschild geometry we take $Q = 0$ to get
\be
 \ve R^2 \lh 1 - u^{t\,2} \rh =  R(R - 3M) u^t - \lh R^2 - 3MR + 3M^2 \rh u^{t\,3}.
\label{5.9}
\ee
It is straightforward to check that in the spinless case with
\be
u^t= \frac{\ve}{1 - \frac{2M}{R}},
\label{5.10}
\ee
this relation reduces to the standard result
\be
\ve = \frac{1 - \frac{2M}{R}}{\sqrt{1 - \frac{3M}{R}}},
\label{5.11}
\ee
for circular orbits. 

Having solved eq.\ (\ref{5.8}) ---or (\ref{5.9}) if appropriate--- one can then also compute the 
orbital angular momentum per unit of mass from the hamiltonian constraint:
\be
\ell^2 = R^4 u^{\vf\,2} =  \lh R^2 - 2MR + Q^2 \rh u^{t\,2} - R^2.
\label{5.12}
\ee
For neutral particles ($q = 0$) instead of eliminating $\eta$ and $u^{\vf}$ from eqs.\ (\ref{5.2}) and (\ref{5.7}), 
one can alternatively eliminate $\ve$ and $u^t$ to get a relation for the total angular momentum per 
unit of mass $\eta$ and the proper angular velocity $u^{\vf}$:
\be
\ba{l} 
\eta \lh MR - Q^2 \rh^2 \left[ 2 MR - 3Q^2 + R^2 u^{\vf\,2} \lh R^2 - 2Q^2 \rh \right]  \\
 \\
 \hs{2} =\, R^2 u^{\vf} \lh M^2R^4 - 2M^2R^2 Q^2 - 2MR^3 Q^2 + R^2 Q^4 + 4MRQ^4 - 2Q^6 \rh \\
 \\
 \hs{3.3} -\, R^4 u^{\vf\,3} \lh MR^5 - 6 M^2 R^4 + 6 M^3R^3 - R^4Q^2 + 10 MR^3Q^2 - 11 M^2R^2Q^2 \rd \\
 \\
 \hs{7.5} \ld -\, 4R^2Q^4 + 5 MRQ^4 \rh.
\ea
\label{5.12.1}
\ee
In Schwarzschild space-time this reduces to the result quoted in ref.\ \ct{d'Ambrosi:2015gsa}:
\be
\frac{\eta}{M} \lh 2 + \frac{R^3u^{\vf\,2}}{M} \rh = 
 \frac{R^3 u^{\vf}}{M^2} \left[ 1 - \frac{R^3u^{\vf\,2}}{M} \lh 1 - \frac{6M}{R} + \frac{6M^2}{R^2} \rh \right]. 
\label{5.12.2}
\ee

\section{Plane non-circular orbits \label{ps6}}

In General Relativity the standard procedure for comparing geodesics is the method of geodesic deviations. 
It is based on a covariant definition of differences between geometric quantities associated with geodesics,
like the unit tangent vectors defining the proper four-velocities of test particles. The procedure can be generalized
to world-lines of particles carrying charge and/or spin as follows. 

Consider two solutions $(x^{\mu}(\tau), u^{\mu}(\tau), \Sg^{\mu\nu}(\tau))$ and 
$(\bar{x}^{\mu}(\tau), \bar{u}^{\mu}(\tau), \bar{\Sg}^{\mu\nu}(\tau))$ of the equations of motion (\ref{2.6}) 
and (\ref{2.7}). The direct differences between dynamical quantities on each world-line at equal proper time
$\tau$ are denoted by $\del$:
\be
\del X(\tau) = \bar{X}(\tau) - X(\tau),
\label{c6.0}
\ee
for any $X = (x^{\mu}, u^{\mu}, \Sg^{\mu\nu})$. As the co-ordinates $x^{\mu}$ are space-time scalars, 
the velocities $u^{\mu}$ space-time vectors and the spin-dipoles $\Sg^{\mu\nu}$ space-time tensors, 
we can define their covariant differences by parallel displacement: 
\be
\ba{l}
\Del x^{\mu}(\tau) = \del x^{\mu}(\tau), \hs{1} 
\Del u^{\mu} = \del u^{\mu}(\tau) + \del x^{\lb} \Gam_{\lb\nu}^{\;\;\;\mu} u^{\nu}, \\
 \\
\Del \Sg^{\mu\nu} = \del \Sg^{\mu\nu} + \del x^{\lb} \Gam_{\lb\kg}^{\;\;\;\mu} \Sg^{\kg\nu} 
 + \del x^{\lb} \Gam_{\lb\kg}^{\;\;\;\nu} \Sg^{\mu\kg}.
\ea
\label{c6.1}
\ee
The equations of motion now imply equations for the proper-time dependence of these covariant 
variations; to linear order: 
\be
\ba{l}
\Del u^{\mu} = D_{\tau} \Del x^{\mu}, \\
 \\
\dsp{ D_{\tau}^2 \Del x^{\mu} - R_{\lb\kg\;\,\nu}^{\;\;\;\,\mu} u^{\kg} u^{\nu} \Del x^{\lb} = 
 \frac{1}{2m}\, \Sg^{\rg\sg} R_{\rg\sg\;\,\nu}^{\;\;\;\,\mu}\, D_{\tau} \Del x^{\nu} + 
 \frac{1}{2m}\, \Sg^{\rg\sg} \nb_{\lb} R_{\rg\sg\;\,\nu}^{\;\;\;\,\mu} u^{\nu} \Del x^{\lb} }\\
 \\
\dsp{ \hs{12} +\, \frac{1}{2m}\, \Del \Sg^{\rg\sg} R_{\rg\sg\;\,\nu}^{\;\;\;\,\mu} u^{\nu} + \frac{q}{m}\, 
  F^{\mu}_{\;\,\nu} D_{\tau} \Del x^{\nu} + \frac{q}{m}\, \nb_{\lb} F^{\mu}_{\;\,\nu}\, u^{\nu} \Del x^{\lb}, }\\
 \\
D_{\tau} \Del \Sg^{\mu\nu} + \lh R_{\lb\kg\sg}^{\;\;\;\;\;\,\mu} \Sg^{\sg\nu} 
 - R_{\lb\kg\sg}^{\;\;\;\;\;\,\nu} \Sg^{\sg \mu} \rh u^{\kg} \Del x^{\lb} = 0.
\ea
\label{c6.2}
\ee
The formalism, eventually with higher-order extensions 
\ct{Koekoek:2010pv,Koekoek:2011mm,Kerner:2001cw}, can be applied to a 
perturbative construction of world-lines starting from a known solution of the equations of motion. 
The circular orbits found in the previous section define such a starting point to construct eccentric 
planar or non-planar bound orbits in Schwarzschild-Reissner-Norstr{\o}m backgrounds. However, 
computationally it is simpler to work with the non-covariant variations (\ref{c6.0}) rather than the 
covariant ones (\ref{c6.1}).

Considering eccentric orbits in the equatorial plane we keep the conditions (\ref{5.1}, \ref{5.0}). With 
these restrictions the allowed deviations from circular orbits can be parametrised by 
$\del x^{\mu} = (\del t, \del r, \del \vf)$ for the orbital and 
$\del \Sg^{\mu\nu} = (\del \Sg^{tr}, \del \Sg^{r\vf}, \del \Sg^{t\vf})$ for the spin degrees of freedom. 
Now using the conservation laws the variations $\del \Sg^{tr}$ and $\del \Sg^{r\vf}$ can equivalently 
be expressed by the change in energy $\del \ve$ and total angular momentum $\del \eta$ and the 
co-ordinate variations. In a condensed notation the relevant linearised deviation equations (\ref{c6.2}) 
then reduce to
\vs{1}

\be
\ba{l}
\lh \ba{cccc}
 \frac{d^2}{d\tau^2} & 0 & \ag \frac{d}{d\tau} & \bg \\
  & & & \\
  0 & \frac{d^2}{d\tau^2} & \gam \frac{d}{d\tau} & \zg \\
  & & & \\
  \kg \frac{d}{d\tau} & \lb \frac{d}{d\tau} & \frac{d^2}{d\tau^2} + \mu & 0 \\
  & & & \\
  \nu \frac{d}{d\tau} & \sg \frac{d}{d\tau} & \chi & \frac{d}{d\tau}
 \ea \rh \lh \ba{c} \del t \\ \\ \del \vf \\ \\ \del r \\ \\ \del \Sg^{t\vf} \ea \rh 
  = \lh \ba{c} 0 \\ \\  0 \\ \\ a \del \eta + b \del \ve \\  \\ c \del \eta + d \del \ve \ea \rh,
\ea
\label{6.1}
\ee
\vs{1}

\nit
where the coefficients are defined in terms of the parameters of the circular reference orbit; the
explicit expressions are given in appendix \ref{ac}.

The general solution of the inhomogenous linear equations (\ref{6.1}) can be decomposed in a specific solution 
plus a solution of the homogeneous equation. Now it is easy to find a simple specific solution: a constant shift 
$\del r$ such that the new circular orbit has the same energy $\ve' = \ve + \del \ve$ and total angular momentum 
$\eta' = \eta + \del \eta$ as the non-circular orbit we wish to construct. Hence by taking this special circular orbit 
as the reference orbit we fix $\del \ve = \del \eta = 0$, and we only have to solve the homogeneous equation (\ref{6.1}). 

To solve the homogeneous equation, we consider the characteristic equation for the periodic eigenfunctions of 
the operator (\ref{6.1}):
\be
\og^3 \lh \og^4 - A \og^2 + B \rh = 0, 
\label{6.5}
\ee
where
\be
\ba{l}
A = \mu - \ag \kg - \bg \nu - \gam \lb - \zg \sg, \\
 \\
B = \bg \lh \kg \chi - \mu\nu + \gam (\lb \nu - \kg \sg) \rh + \zg \lh \lb \chi - \mu \sg - \ag (\lb \nu - \kg \sg) \rh.
\ea
\label{6.6}
\ee
In addition to three 0-modes for potential secular solutions familiar from the motion of spinless particles
\ct{Koekoek:2010pv,Koekoek:2011mm} there are two pairs of non-trivial periodic solutions 
with angular frequencies
\be
\og^2_{\pm} = \frac{1}{2} \lh A \pm \sqrt{A^2 - 4B} \rh.
\label{6.7}
\ee
These periodic solutions can be cast in the simple form 
\be
\ba{l}
\del t = n^t_+ \sin \og_+ (\tau - \tau_+) + n^t_- \sin \og_- (\tau - \tau_-), \\
 \\
\del \vf = n^{\vf}_+  \sin \og_+ (\tau - \tau_+) + n^{\vf}_- \sin \og_- (\tau - \tau_-), \\ 
 \\
\del r = n^r_+ \cos \og_+ (\tau - \tau_+) + n^r_- \cos \og_- (\tau - \tau_-),\\
 \\ 
\del \Sg^{t\vf} = n^{\sg}_+ \sin \og_+ (\tau - \tau_+) + n^{\sg}_- \sin \og_- (\tau - \tau_-), 
\ea
\label{6.8}
\ee
where $\tau_{\pm}$ are constants of integration determining the relative phases of the oscillations, 
and up to some common normalization constants $C_{\pm}$ the amplitudes are 
\be
\ba{l} 
n^t_{\pm} =  C_{\pm} \left[ \lb \lh \bg \gam - \ag \zg \rh + \bg \lh \og_{\pm}^2 - \mu \rh \right], \\
 \\
n^{\vf}_{\pm} = C_{\pm} \left[ - \kg \lh \bg \gam - \ag \zg \rh + \zg \lh \og^2_{\pm} - \mu \rh \right], \\
 \\
n^r_{\pm} = C_{\pm}\, \og_{\pm} \lh \bg \kg + \zg \lb \rh, \\
 \\
n^{\sg}_{\pm} = C_{\pm}\, \og^2_{\pm} \lh \og^2_{\pm} - \mu + \ag \kg + \gam \lb \rh. 
\ea
\label{6.9}
\ee
The null solutions of eq.\ (\ref{6.5}) suggest that in addition to the periodic solutions (\ref{6.8}) 
there might also be secular solutions for the orbital degrees of freedom \ct{Koekoek:2010pv}. 
However, as we have chosen the energy and total angular momentum of the orbit to be the same 
as that of the circular reference orbit no such freedom is left in this case, except for trivial shifts
in the origin of the $t$- and $\vf$-co-ordinates. Therefore the complete first-order solution for the 
non-circular planar orbits is
\be
\ba{l}
t(\tau) = u^t \tau + n^t_+ \sin \og_+ (\tau - \tau_+) + n^t_- \sin \og_- (\tau - \tau_-), \\
 \\
\vf(\tau) = u^{\vf} \tau + n^{\vf}_+  \sin \og_+ (\tau - \tau_+) + n^{\vf}_- \sin \og_- (\tau - \tau_-), \\ 
 \\
r(\tau) = R + n^r_+ \cos \og_+ (\tau - \tau_+) + n^r_- \cos \og_- (\tau - \tau_-),\\
 \\ 
\Sg^{t\vf}(\tau) = n^{\sg}_+ \sin \og_+ (\tau - \tau_+) + n^{\sg}_- \sin \og_- (\tau - \tau_-).
\ea
\label{6.10}
\ee
The perturbations, in particular those in the radial direction, have double periods. Hence the periastron 
and apastron will behave in a complicated way, as the body reaches different minimal or maximal 
radial distances at non-constant intervals. However, in the limit $B \ll A^2$ the dominant frequency
will be $\og_+ \simeq \sqrt{A}$, and the variations in the periastron and apastron will be relatively slow.
An example for the case of Schwarzschild geometry $(Q = 0)$ is given in figure 1, where we have plotted 
the radial variation as a function of proper time for a circular reference orbit $R = 10 M$ with orbital angular 
momentum $\ell = 4M$ and for deviation parameters 
\be
n_+^r = 0.1 R, \hs{1} n_-^r = 0.05 R, \hs{1} \tau_+ - \tau_- = 100 M.
\label{6.10.a}
\ee

\bc
\scalebox{0.5}{\hs{4} \includegraphics{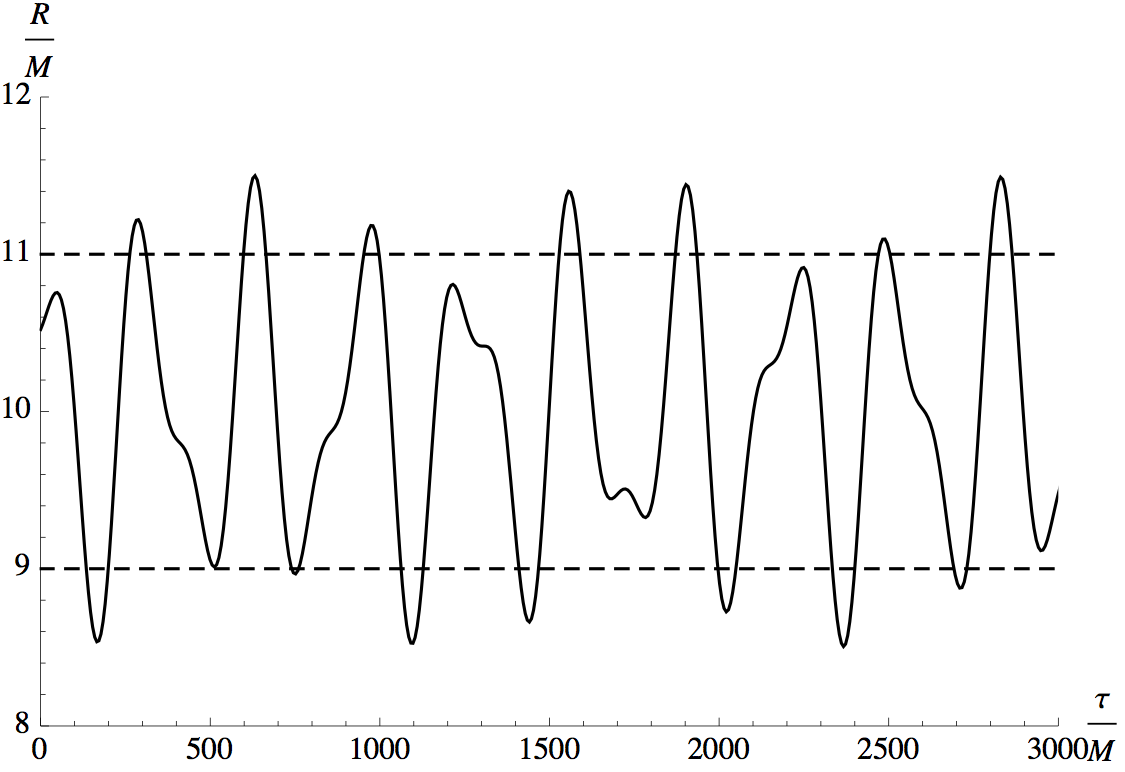}}
\vs{2}

{\footnotesize \begin{tabular}{lc} Fig.\ 1: & Radial deviation from circular orbit  with $R = 10 M$ and $\ell = 4M$ \\
 & as a function of proper time in Schwarzschild space-time for \\ & deviation parameters as in eq.\ (\ref{6.10.a}). \end{tabular}}
\ec

\nit
As we have obtained the non-circular orbits (\ref{6.10}) under the conditions that the specific energy and
total angular momentum are the same as those of the circular reference orbits, the conservation laws 
(\ref{4.3}) and (\ref{4.5}) for the spin-dipole components link the variations $\del \Sg^{\mu\nu}$ of these 
quantities to those of the orbital parameters
\be
\del \Sg^{r\vf} = F \del u^{\vf} + G\, \frac{\del r}{R}, \hs{2}
\del \Sg^{tr} =  K \del u^t + L\, \frac{\del r}{R}, 
\label{6.13}
\ee
where
\be
\ba{l}
\dsp{ F = - mR, \hs{2} G = - \frac{m}{R} \lh \eta + R^2 u^{\vf} \rh, }\\
 \\
\dsp{ K = \frac{mR^3}{MR - Q^2} \lh 1 - \frac{2M}{R} + \frac{Q^2}{R^2} \rh, }\\
 \\
\dsp{ L =  \frac{mR^4}{(MR - Q^2)^2} \left[ \lh \frac{2M}{R} \lh 1 - \frac{M}{R} + \frac{2Q^2}{R^2} \rh 
  - \frac{Q^2}{R^2} \lh 3 + \frac{Q^2}{R^2} \rh \rh u^t \rd }\\
 \\
\dsp{ \hs{2}  \ld -\, \lh \frac{2M}{R} - \frac{3Q^2}{R^2} \rh \ve 
  + \frac{qQ}{mR} \lh \frac{M}{R} - \frac{2Q^2}{R^2} \rh \right]. }
\ea
\label{6.13.1}
\ee
As a result
\be
\ba{l}
\del \Sg^{r\vf} = N_+^{r\vf} \cos \og_+ (\tau - \tau_+) + N_-^{r\vf} \cos \og_- (\tau - \tau_-), \\
 \\
\del \Sg^{tr} = N_+^{tr} \cos \og_+ (\tau - \tau_+) + N_-^{tr} \cos \og_- (\tau - \tau_-), 
\ea
\label{6.14}
\ee
with
\be
N^{r\vf}_{\pm} = \og_{\pm} F n^{\vf}_{\pm} + G\, \frac{n^r_{\pm}}{R}, \hs{2} 
N^{tr}_{\pm} =  \og_{\pm} K n^t_{\pm} + L\, \frac{n^r_{\pm}}{R}. 
\label{6.15}
\ee
Thus we have obtained a large class of non-circular planar orbits (to first order in the deviations), parametrised by 
the constants $C_{\pm}$ and the radial co-ordinate $R$ of the circular orbit with the same specific energy and 
total angular momentum.

\section{The ISCO \label{4pisco}}

In black-hole space-times there is an innermost stable circular orbit (ISCO) at a specific distance 
from the horizon. For simple point masses in Schwarzschild space-time this orbit is located at $R = 6M$.  
Here the effective potential has a flex point where the orbital angular momentum reaches a minimum of 
$\ell^2 = 12 M^2$. Circular orbits at values $R < 6M$ are possible in principle but not in practice, as 
they correspond to maxima of the effective potential rather than minima; thus they are unstable 
under small perturbations. 

The presence of spin alters the stability conditions and therefore the location of the ISCO. The stability 
conditions for circular orbits can be derived directly from the analysis in the previous section. Indeed, 
circular orbits are stable as long as the planar deviations display oscillatory behaviour. In contrast, 
whenever the frequency $\og$ of these deviations develops an imaginary part the radial motion displays 
exponential behaviour and the orbit becomes unstable \ct{Suzuki:1998isco,Blanchet:2002mb}. Now the 
frequencies of the deviations are solutions of the eigenvalue equation (\ref{6.7}). Thus we must ask what 
is the parameter domain in which the eigenvalues are real, and especially where the boundary between 
stability and instability is located. The first condition is obviously for $\og_{\pm}^2$ to be real; this requires
\be
A^2 - 4B \geq 0.
\label{isco.1}
\ee
In addition, for the frequencies $\og_{\pm}$ themselves to be real as well we must demand that 
$\og_{\pm}^2 \geq 0$, which happens if 
\be
A \geq 0, \hs{2} B \geq 0.
\label{isco.2}
\ee
Fig.\ 2 shows the solutions of these inequalities for the case of Schwarzschild space-time in terms 
of the allowed values of the dimensionless radial co-ordinate $R/M$ and of the orbital angular 
momentum per unit of mass
\be
\ell = R^2 u^{\vf}.
\label{isco.3}
\ee
The shaded area corresponds to stable circular orbits. As we have established in sect.\ \ref{ps5} that 
any circular orbit is determined for a given background geometry by the parameters $R$ and $J$ 
---which fix $u^{\vf}$, $u^t$ and $E$--- the allowed orbits for fixed $R/M$ and various $\ell/M$ in 
figure 2 differ in the values of $\eta = J/m$. Equivalently they differ in the value of the spin per unit 
of mass parametrized by the dimensionless variable
\be
\frac{\sg}{M} = \frac{R\Sg^{r\vf}}{mM}. 
\label{isco.4}
\ee
\vs{1}

\bc
\scalebox{0.95}{\includegraphics{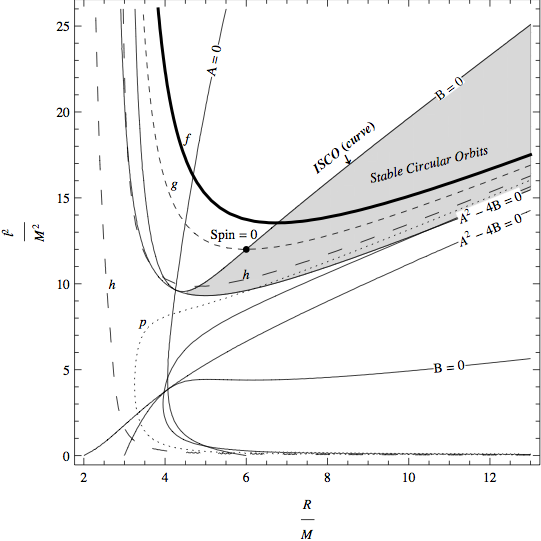}}

\vs{1}
{\footnotesize \begin{tabular}{lc} Fig.\ 2: & 
 Allowed domains of radius $R/M$ and orbital angular momentum $\ell/M$ for plane \\
 & circular  orbits in Schwarzschild space-time. Included are four curves labeled \\ 
 & {\em f, g, h, p} defining orbits of fixed spin per unit of mass $\sg$ for retrograde, \\ 
 & vanishing and prograde spin $ \sg/M = (-0.5, 0, 0.5, 0.7)$. \end{tabular}}
\ec

\nit
We have also indicated in figure 2 several curves of constant spin. The curves {\em g, f, h} 
represent iso-spin lines for spin $\sg = 0$, for retrograde spin $\sg  = - 0.5M$ and for prograde spin
$\sg = 0.5M$, respectively. In each case the ISCO is defined by the value of $R/M$ where the 
curve crosses the line $B = 0$. For vanishing spin this is at the well-known value $R = 6M$, for 
prograde spin it is at a lower value of $R$ and for retrograde spin at a higher value of $R$. There
actually is a smallest ISCO $R \simeq 4.5 M$ for $\sg \simeq 0.55 M$, where the curve $B=0$ 
reaches a minimum value of orbital angular momentum. 

For higher spin values the ISCO is reached at the point where the iso-spin curves cross the line 
$A^2 - 4B = 0$; an example is the curve labeled {\em p} corresponding to $\sg = 0.7 M$. In 
general such high values of $\sg/M$ are possible only if the masses $m$ and $M$ of the test 
particle and the black hole creating the background become comparable. Of course the back 
reaction of the test mass can then no longer be ignored and our estimates of the ISCO 
become unreliable. 

By calculating the values of the spin parameter $\sg/M$ on the lines separating regions of stable 
and unstable orbits we have extracted the values of $R$ for the ISCO as a function of $\sg/M$ in 
Schwarzschild space-time. The result is represented by the continuous curve labeled $R_{isco}$ 
in figure 3. The two branches correspond to lower-spin ISCOs on the curve $B = 0$ and 
higher-spin ISCOs on the curve $A^2 - 4B = 0$, respectively. These results agree qualitatively 
with other studies in the literature based on the conventional Mathisson-Papapetrou-Dixon 
approach \ct{Suzuki:1998isco,Hackmann:2014}. 

The iso-spin curves in figure 2 for lower-spin values, corresponding to the left-hand branch in 
figure 3, also suggest that the circular orbits become unstable when the orbital angular momentum 
reaches its minimum as a function of radial distance for contant $\sg$. 
\vs{3}

\bc
\scalebox{1}{\includegraphics{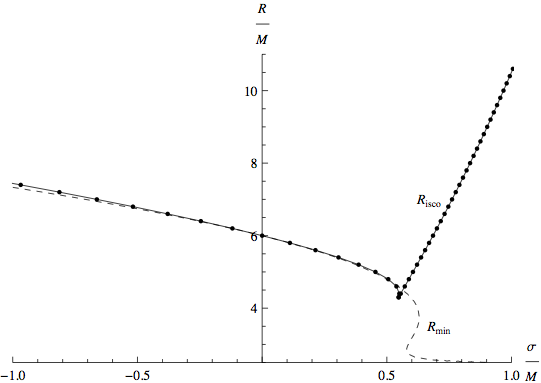}}
\vs{2}

{\footnotesize  \begin{tabular}{rc} Fig.\ 3:  & Radius of the ISCO: $R_{isco}/M$, as a function of spin 
   $\sg/M$ for Schwarzschild space-time \\ & (continuous curve)  compared with the radius of minimal 
   orbital angular momentum\\ & $R_{\min}/M$ at fixed spin (dashed curve). \end{tabular} }
\ec

\np
\nit
This issue can be analysed more precisely by returning to eq.\ (\ref{5.12.2}) and rewriting it in the form 
\be
\frac{\sg R}{M^2} \lh \frac{2R}{M} + \frac{\ell^2}{M^2} \rh = 
 \frac{\ell}{M} \lh \frac{R}{M} - 2 \rh \left[ \frac{R^2}{M^2} - \frac{\ell^2}{M^2} \lh \frac{R}{M} - 3 \rh \right].
\label{isco.5}
\ee
From this equation one can derive the minimum of $\ell/M$ as a function of $R/M$ for fixed spin 
$\sg$. The result is plotted as the dashed curve labeled $R_{min}$ in figure 3. Over the range of 
predominant physical interest $-0.5 < \sg/M < 0.5$ the curve nearly coincides with that for $R_{isco}$. 
However there are small differences for larger absolute spin values, reminiscent of those found in 
higher-order PN-corrections for compact binaries \ct{Blanchet:2002mb}. Note that the parts of the 
dashed curve for large retrograde spin actually enter the region of instability, hence do not correspond 
to stable circular orbits.
\vs{2}

\bc
\scalebox{01}{\includegraphics{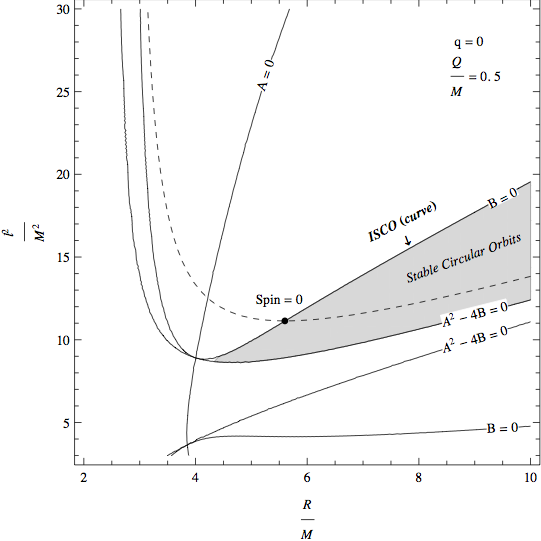}}
\vs{1}

{\footnotesize  \begin{tabular}{rc} Fig.\ 4:  & Stable circular orbits of a chargeless spinning
particle in \\ & Reissner-Nordstr{\o}m space-time with $Q/M = 0.5$. \end{tabular} }
\ec
\vs{1}

\nit
A similar analysis can be done for the case of a spinning test-particle in a Reissner-Nordstr{\o}m 
background. Of course one has two more parameters to account for: the charge $Q$ of the 
central black hole and the charge $q$ of the test particle. Figure 4 shows the stability region of 
circular orbits of electrically neutral spinning test particles for a black hole of charge $Q = 0.5 M$,
including the curve representing the $\ell$-$R$-relation for circular orbits of a spin-0 particle. In 
figure 5 the radial co-ordinate of the ISCO is given as a function of spin for the same black-hole 
background geometry with $Q = 0.5 M$. We observe that the ISCOs are located closer to the 
horizon: $R_{hor} \simeq 1.87 M$, and that with this value of the black-hole charge the minimal 
ISCO is reached at a somewhat smaller spin value than in the Schwarzschild case: $\sg_{min} 
\simeq 0.53 M$. 
\vs{1}

\bc
\scalebox{1}{\includegraphics{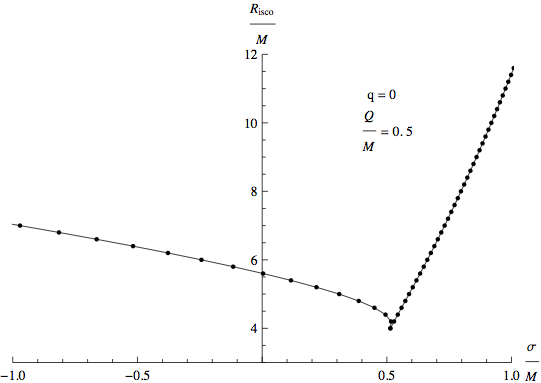}}
\vs{1}

{\footnotesize  \begin{tabular}{rc} Fig.\ 5:  & $R_{isco}/M$ as a function of $\sg/M$ for 
   Reissner-Norstr{\o}m space-time with $Q/M = 0.5$. \end{tabular} }
\ec

\section{A non-minimal hamiltonian \label{ps7}}

In the presence of external fields spinning particles can be subject to spin-dependent forces coupling 
to gradients in the fields like the well-known Stern-Gerlach force 
\ct{Barducci:1976qu,Khriplovich:1989ed,vanHolten:1987pro,vanHolten:1990we} in electrodynamics. 
Such forces can be modeled in our approach by additional spin-dependent terms in the hamiltonian:
\be
H = H_0 + H_{SG}, \hs{2} 
H_{SG} = \frac{\kg}{4}\, R_{\mu\nu\kg\lb} \Sg^{\mu\nu} \Sg^{\kg\lb} + \frac{\lb}{2}\, F_{\mu\nu} \Sg^{\mu\nu}.
\label{7.1}
\ee
Using this non-minimal hamiltonian in the brackets to construct equations of motion we get
\be
\ba{l}
\pi_{\mu} = m g_{\mu\nu} u^{\nu}, \\
 \\
\dsp{ m g_{\mu\nu} D_{\tau} u^{\nu} = \frac{1}{2}\, \Sg^{\kg\lb} R_{\kg\lb\mu\nu} u^{\nu} + q F_{\mu\nu} u^{\nu} 
 - \frac{\kg}{4}\, \Sg^{\rg\sg} \Sg^{\kg\lb}\, \nb_{\mu} R_{\rg\sg\kg\lb} - \frac{\lb}{2}\, \Sg^{\kg\lb}\, \nb_{\mu} F_{\kg\lb}, }
\ea
\label{7.2}
\ee
and
\be
D_{\tau} \Sg^{\mu\nu} = \lh \kg \Sg^{\rg\sg} R_{\rg\sg\;\,\lb}^{\;\;\;\,\mu} + \lb F^{\mu}_{\;\,\lb} \rh \Sg^{\nu\lb}
 - \lh \kg \Sg^{\rg\sg} R_{\rg\sg\;\,\lb}^{\;\;\;\,\nu} + \lb F^{\nu}_{\;\,\lb} \rh \Sg^{\mu\lb}.
\label{7.3}
\ee
Remarkably, using eqs.\ (\ref{3.6}, \ref{3.7}) and the Bianchi identities for $F_{\mu\nu}$ and $R_{\mu\nu\kg\lb}$
it is straightforward to generalize the theorem of ref. \ct{d'Ambrosi:2015gsa}, that any constant of motion (\ref{3.4}) 
remains a constant of motion in the presence of Stern-Gerlach forces:
\be 
\ba{lll}
\left\{ J, H_{SG} \right\} & = & \dsp{ \kg\, \Sg^{\mu\nu} \Sg^{\rg\sg} \lh - \frac{1}{4}\, \ag^{\lb} \nb_{\lb} R_{\rg\sg\mu\nu}
 + R_{\rg\sg\mu}^{\;\;\;\;\;\,\lb} \bg_{\lb\nu} \rh }\\
 & & \\
 & & \dsp{ + \lb\, \Sg^{\mu\nu} \lh - \frac{1}{2}\, \ag^{\lb} \nb_{\lb} F_{\mu\nu} 
 + F_{\mu}^{\;\,\lb} \bg_{\lb\nu} \rh = 0. }
\ea
\label{7.4}
\ee
In particular in Schwarzschild and Reissner-Nordstr{\o}m backgrounds the energy $E$, eq.\ (\ref{4.3}), and all 
three components of the full total angular momentum $\bfJ$, eqs.\ (\ref{4.4}), remain constants of motion for 
spinning neutral and charged particles. 

The existence of these constants of motion enable us to perform a similar analysis of orbits as in the case of 
the minimal hamiltonian. As an example we solve for plane circular orbits in a Schwarzschild  background 
similar to those discussed in section \ref{ps5} for the minimal case. Again we take the plane of the orbit to be 
the equatorial plane $\thg = \pi/2$ with $u^{\thg} = u^r = 0$ and $\Sg^{\thg\nu} = 0$ for all values of $\nu$. Then the 
conservation laws reduce to the same form as in eqs.\ (\ref{4.3}) and (\ref{4.5}):
\[
E = m \lh 1 - \frac{2M}{R} \rh u^t - \frac{M}{R^2}\, \Sg^{tr}, \hs{1}
J = m R^2 u^{\vf} + R \Sg^{r\vf}.
\]
The hamiltonian constraint now reads 
\be
H = H_0 + H_{SG} = - \frac{m}{2},
\label{7.5}
\ee
which becomes 
\be
- \lh 1 - \frac{2M}{R} \rh u^{t\,2} + R^2 u^{\vf\,2} + 1 = - \frac{2\kg M}{m} \left[ \frac{2}{R^3}\, \Sg^{tr\,2} 
 + \frac{\Sg^{r\vf\,2}}{R - 2M} - \frac{1}{R} \lh 1 - \frac{2M}{R} \rh \Sg^{t\vf\,2} \right].
\label{7.6}
\ee
Next the total spin is 
\be
I = - \Sg^{tr\,2} - R^2 \lh 1 - \frac{2M}{R} \rh \Sg^{t\vf\,2} + \frac{R^2 \Sg^{r\vf\,2}}{1 - \frac{2M}{R}}, 
\label{7.6.1}
\ee
a constant. These constraints plus the vanishing of the radial acceleration: $\dot{u}^r = 0$ imply that  
the angular velocity and the time-dilation factor are constant, causing in fact $\Sg^{t\vf}$ to vanish:
\be
\dot{u}^t = \dot{u}^{\vf} = \Sg^{t\vf} = 0.
\label{7.7}
\ee
Then one finds a quartic equation for the angular velocity in terms of the radius $R$ and the angular momentum 
$J = m \eta$, which generalizes eq.\ (\ref{5.12.2}):
\be
\ba{l} 
\dsp{ \lh 1 - \frac{2M}{R} \rh^2 R u^{\vf} \left[ \frac{2M^2 \eta}{R^3} - M u^{\vf} + M \eta u^{\vf\,2} 
 + \lh R^2 - 6 MR + 6 M^2 \rh R u^{\vf\,3} \right] }\\
 \\
\dsp{ \hs{7} =\, \kg m A + \kg^2 m^2 B + \kg^3 m^3 C,}
\ea
\label{7.8}
\ee
where the coefficients $(A, B, C)$ are quartic polynomials in $u^{\vf}$ themselves; the explicit expressions
are given in appendix \ref{ab}. After solving this equation for $u^{\vf}$ also the energy $E = m \ve$ and the 
values of $u^t$, $\Sg^{tr}$ and $\Sg^{r\vf}$ can be determined.

\section{Energy-momentum tensor \label{ps8}} 

In the previous sections the equations of motion for a relativistic spinning particle were obtained starting from 
a closed set of brackets (\ref{2.4}) and the choice of a hamiltonian. The same equations can be derived by 
energy-momentum conservation using an appropriate energy-momentum tensor \ct{Steinhoff:2009tk,vanHolten:2015vfa}. This tensor then also defines the source term in the Einstein equations to compute the back reaction of the particle on the 
space-time geometry; indeed, the Einstein equations require the energy momentum to be divergence-free
\be
G_{\mu\nu} = R_{\mu\nu} - \frac{1}{2}\, g_{\mu\nu} R = - 8 \pi G T_{\mu\nu} \hs{1} \Rightarrow \hs{1}
\nb^{\mu} G_{\mu\nu} = - 8 \pi \nb^{\mu} T_{\mu\nu} = 0.
\label{8.1}
\ee
This identity is to be guaranteed by the equations of motion. For a neutral particle described by the minimal 
hamiltonian this follows by taking\footnote{We define the delta-function as a scalar density of weight 1/2, 
such that for scalar functions $f(x)$
\[
\int d^4y\, \del^4(x - y) f(y) = f(x).
\label{8.3}
\]}
\be
T_0^{\mu\nu} = m \int d\tau\, u^{\mu} u^{\nu}\, \frac{1}{\sqrt{-g}}\, \del^4\lh x - X \rh
 + \frac{1}{2}\, \nb_{\lb}  \int d\tau \lh u^{\mu} \Sg^{\nu\lb} + u^{\nu} \Sg^{\mu\lb} \rh \frac{1}{\sqrt{-g}}\, \del^4\lh x - X \rh.
\label{8.2}
\ee
The covariant divergence of $T_0^{\mu\nu}$ is
\be
\ba{lll}
\nb_{\mu} T_0^{\mu\nu} & = & \dsp{ \int d\tau \lh m \frac{D u^{\nu}}{D\tau} -  
 \frac{1}{2}\, \Sg^{\kg\lb} R_{\kg\lb\;\mu}^{\;\;\;\,\nu} u^{\mu} \rh\frac{1}{\sqrt{-g}}\, \del^4\lh x - X \rh }\\
 & & \\
 & & \dsp{ +\, \frac{1}{2}\, \nb_{\lb} \int d\tau \frac{D\Sg^{\nu\lb}}{D\tau}\, \frac{1}{\sqrt{-g}}\, \del^4\lh x - X \rh = 0.}
\ea
\label{8.4}
\ee
and vanishes upon applying the equations of motion (\ref{2.6}, \ref{2.7}) with $q = 0$. Similarly, for a particle 
subject to the gravitational Stern-Gerlach force with the hamiltonian $H_0 + H_{SG}$ the correct expressions is
\be
T^{\mu\nu} = T_0^{\mu\nu} + \kg\, T_1^{\mu\nu}, 
\label{8.5}
\ee
where
\be
\ba{lll}
T_1^{\mu\nu} & = & \dsp{ \frac{1}{2}\, \nb_{\kg} \nb_{\lb} \int d\tau \lh \Sg^{\mu\lb} \Sg^{\kg\nu} 
 + \Sg^{\nu\lb} \Sg^{\kg\mu} \rh \frac{1}{\sqrt{-g}}\, \del^4\lh x - X \rh }\\
 & & \\
 & & \dsp{ +\, \frac{1}{4} \int d\tau\, \Sg^{\rg\sg} \lh R_{\rg\sg\lb}^{\;\;\;\;\;\,\nu} \Sg^{\lb\mu}
  + R_{\rg\sg\lb}^{\;\;\;\;\;\,\mu} \Sg^{\lb\nu} \rh \frac{1}{\sqrt{-g}}\, \del^4\lh x - X \rh. }
\ea
\label{8.6}
\ee
Again performing standard operations from tensor calculus including Ricci- and Bianchi-identities 
leads to the result
\be
\ba{lll}
\nb_{\mu} T_1^{\mu\nu} & = & \dsp{ \frac{1}{4} \int d\tau\, \nb^{\nu} R_{\rg\sg\kg\lb}\, \Sg^{\rg\sg} \Sg^{\kg\lb}
 \frac{1}{\sqrt{-g}}\, \del^4\lh x - X \rh }\\
 & & \\
 & & \dsp{ +\, \frac{1}{2}\, \nb_{\lb} \int d\tau\, \Sg^{\rg\sg} \lh R_{\rg\sg\kg}^{\;\;\;\;\;\,\lb} \Sg^{\kg\nu}
  - R_{\rg\sg\kg}^{\;\;\;\;\;\,\nu} \Sg^{\kg\lb} \rh \frac{1}{\sqrt{-g}}\, \del^4\lh x - X \rh. }
\ea
\label{8.7}
\ee
Combining this with the expression (\ref{8.4}) for $\nb_{\mu} T_0^{\mu\nu}$ it follows that the divergence of the 
full energy-momentum tensor vanishes 
\be
\nb_{\mu} \lh T_0^{\mu\nu} + \kg T_1^{\mu\nu} \rh = 0, 
\label{8.8}
\ee
provided the non-minimal equations of motion (\ref{7.2}, \ref{7.3}) hold.  Finally, one can also take into account 
the electro-magnetic Lorentz- and Stern-Gerlach forces by additional contributions  
\be
T^{em}_{\mu\nu} = F_{\mu}^{\;\,\lb} F_{\nu\lb} - \frac{1}{4}\, g_{\mu\nu} F_{\kg\lb} F^{\kg\lb} - 
 \frac{\lb}{2}\, g_{\mu\nu} \int d\tau F_{\kg\lb}\, \Sg^{\kg\lb} \frac{1}{\sqrt{-g}}\, \del^4\lh x - X \rh.
\label{8.9}
\ee

\section{Discussion and summary \label{ps9}}

In this paper we have developed a covariant hamiltonian framework for spinning particles in gravitational
and electromagnetic background fields. One of its strong points is that it doesn't require an a priori choice 
of hamiltonian, hence it can be applied to a large variety of models of relativistic spin dynamics. We have 
discussed in particular the case of the minimal kinetic hamiltonian $H_0$ defined in eq.\ (\ref{2.5}) and 
its extension with a Stern-Gerlach type interaction hamiltonian $H_{SG}$ defined in eq.\ (\ref{7.1}). Other 
extensions are possible in principle. For example, one could introduce more covariant tensor fields allowing 
general expressions of the form
\be 
H_{eff} = \frac{1}{2m}\, g^{\mu\nu} \pi_{\mu} \pi_{\nu} + K^{\lb}_{\;\,\mu\nu} \pi_{\lb} \Sg^{\mu\nu} 
 + \frac{1}{2}\, M_{\mu\nu\kg\lb} \Sg^{\mu\nu} \Sg^{\kg\lb},
\label{9.1}
\ee 
where $K^{\lb}_{\;\,\mu\nu}$ and $M_{\mu\nu\kg\lb}$ receive contributions from additional fields and/or 
geometric structures like torsion and curvature. Alternatively the equations of motion for spinning 
particles can be derived from the Einstein equations using appropriate energy-momentum tensors; 
we showed this explicitly for simple spinning particles with minimal or Stern-Gerlach dynamics. 

We have applied our formalism to the case of spinning particles in Schwarzschild and Reissner-Nordstr{\o}m 
backgrounds. We have found interesting new effects. For example the periastron of spinning particles in bound 
orbits is not only subject to an angular shift, but the point of closest approach shows radial variations as well. 
Also the radius of the innermost stable circular orbit changes with spin; over a wide range of spin values 
$-0.5 M < \sg < 0.5 M$ it is quite close to the orbit of minimal orbital 
angular momentum, but only for spinless particles the two actually coincide. In addition we find an
absolute minimal ISCO $R \simeq 4.3M$ at an intermediate point between low and high spin values; 
beyond this the cross-over to instability occurs on a different branch of stability limits, but these values 
are in the regime where the back reaction of the spinning particle on the space-time geometry can no 
longer be neglected. For Extreme Mass Ratio binaries with typical mass ratios $m/M < 10^{-4}$ this 
limit is never reached and the ISCO coincides to good approximation with the circular orbit of minimal
orbital angular momentum.

With the formalism in hand there are still many open problems and applications to be studied. First of all
it would be interesting to consider the effect of spin on the emission of gravitational waves, as well as the 
back reaction of gravitational radiation on the spin dynamics. But also the generalization of our results 
to the case of non-minimal hamiltonians and to Kerr backgrounds with the associated spin-spin coupling 
may have important astrophysical applications. Finally our approach is based on the idealization of compact 
spinning bodies as point-like test particles. It may be refined to apply to finite-size bodies by including 
higher mass multipoles. All of this is left for future investigation.

\vs{2}
\nit
{\bf Acknowledgement} \\
For G.d'A.\ and J.W.v.H.\ this work is part of the research programme {\em Gravitational Physics} of the 
Netherlands Foundation for Fundamental Research on Matter (FOM).
\vs{5}

\np
\appendix
\section{Reissner-Nordstr{\o}m geometry \label{aa}} 

In this appendix we collect the expressions for the components of the connection and Riemann 
curvature tensor used in the main body of the paper. \\
{\em a.\ Connection.\/} From the line element (\ref{4.1}) one derives the following expressions for the 
connection coefficients:
\be
\ba{l}
\dsp{ \Gam_{rt}^{\;\;\;t} = - \Gam_{rr}^{\;\;\;r} = \frac{Mr - Q^2}{r(r^2 - 2Mr + Q^2)},  }\\
 \\
\dsp{  \Gam_{tt}^{\;\;\;r} = \frac{1}{r^5} \lh  Mr - Q^2 \rh \lh r^2 - 2Mr + Q^2 \rh, }\\
 \\
\dsp{ \Gam_{\vf\vf}^{\;\;\;r} = \sin^2 \thg\, \Gam_{\thg\thg}^{\;\;\;r} = - \frac{\sin^2 \thg}{r} \lh r^2 - 2Mr + Q^2 \rh, }\\
 \\
\dsp{ \Gam_{r\thg}^{\;\;\;\thg} = \Gam_{r\vf}^{\;\;\;\vf} = \frac{1}{r}, }\\
 \\
\dsp{ \Gam_{\thg\vf}^{\;\;\;\vf} = \frac{\cos \thg}{\sin \thg}, \hs{2} \Gam_{\vf\vf}^{\;\;\;\thg} = - \sin \thg\, \cos \thg. }
\ea
\label{a1.1}
\ee
{\em b.\ Curvature components.\/} The corresponding curvature two-form 
$R_{\mu\nu} = \frac{1}{2} R_{\kg\lb\mu\nu}\, dx^{\kg} \wedge dx^{\lb}$ has components:
\be
\ba{l}
\dsp{ R_{tr} = \frac{1}{r^4} \lh 2Mr - 3Q^2 \rh dt \wedge dr, }\\ 
 \\
\dsp{ R_{t\thg} = - \frac{1}{r^4} \lh Mr - Q^2 \rh \lh r^2 - 2Mr + Q^2 \rh dt \wedge d\thg, }\\
 \\
\dsp{ R_{t\vf} = - \frac{1}{r^4} \lh Mr - Q^2 \rh \lh r^2 - 2Mr + Q^2 \rh \sin^2 \thg\, dt \wedge d\vf, }\\
 \\
\dsp{ R_{r\thg} = \frac{Mr - Q^2}{r^2 - 2Mr + Q^2}\, dr \wedge d\thg, }\\ 
 \\
\dsp{ R_{r\vf} = \frac{Mr - Q^2}{r^2 - 2Mr + Q^2}\, \sin^2 \thg\, dr \wedge d\vf, }\\
 \\
\dsp{ R_{\thg\vf} = - \lh 2Mr - Q^2 \rh \sin^2 \thg\, d\thg \wedge d\vf. }
\ea
\label{a1.2}
\ee

\np
\section{Coefficients of the deviation equations \label{ac}} 

The deviation equations for planar orbits of spinning particles in Reissner-Nordstr{\o}m space-time w.r.t.\ 
a reference circular orbit $r = R$ can be written in the form (\ref{6.1}). Here we give the expressions for the 
coefficients: 
\be
\ba{ll}
\dsp{ a = \frac{MR - Q^2}{R^3}\, u^{\vf}, }& \dsp{ b = \frac{(2MR - 3 Q^2)(R^2-2MR + Q^2))}{R^3 (MR - Q^2)}\, u^t, }\\
\\
\dsp{ c = - \frac{(MR - Q^2)m u^t }{R^2(R^2 - 2MR + Q^2)}, }& \dsp{ d = \frac{R^2 m u^{\vf}}{MR - Q^2}, }
\ea
\label{ac.1}
\ee
and furthermore
\be
\ba{lll}
\alpha & = & \dsp{ 
  \frac{(2M R^2 (R-M) - R Q^2 (3R-4M) - Q^4)}{R(MR-Q^2)(R^2-2MR+Q^2)}\,  u^t  }\\
 & & \\
  & & \dsp{ - \frac{R(2MR-3Q^2)}{(MR-Q^2)(R^2-2MR+Q^2)}\, \ve + 
         \frac{qQ}{m} \frac{MR - 2Q^2}{(MR - Q^2)(R^2 - 2MR + Q^2)},   }\\
 & & \\
\beta & = & \dsp{ -\frac{(MR-Q^2)}{mR^2}\, u^\vf, }\\
 & & \\
\gamma & = & \dsp{ \frac{2R^2-5MR+3Q^2}{R(R^2-2MR+Q^2)}\, u^\vf + \frac{(MR-Q^2)}{R^3(R^2-2MR+Q^2)}\, \eta, }\\
 & & \\
\zeta & = & \dsp{ -\frac{(R^2-2MR+Q^2)(MR-Q^2)}{mR^6}\, u^t, }\\
 & & \\
\kappa & = & \dsp{ -  \frac{(2MR-3Q^2)(R^2-2MR+Q^2)}{R^3(MR-Q^2)}\, \ve 
 + \frac{qQ}{m} \frac{(MR-2Q^2)(R^2-2MR+Q^2)}{R^4(MR-Q^2)} }\\
 & & \\ 
\lambda & = & \dsp{ \frac{2(MR^2-Q^2(2R-M))}{MR-Q^2}\, u^{\vf} - \frac{MR-Q^2}{R^3}\, \eta, }\\
\ea
\label{ac.2}
\ee
and also 
\[
\ba{lll}
\mu & = & \dsp{ -\frac{2M^2R^3(R-3M)-3MR^2Q^2(2R-7M)+RQ^4(3R-20M)+6Q^6}{R^4(MR-Q^2)^2} }\\
 & & \\
 & & \dsp{ +\, \frac{2M^2R^3(R-4M)-2MR^2 Q^2 (3R - 14M)+RQ^4(3R-28M) + 9Q^6}{R^4(MR-Q^2)^2}\, \ve u^t }\\
 & & \\
 & & \dsp{ - \frac{qQ}{m} \frac{2M^2R^3(R-3M) - MR^2Q^2(7R-24M) + RQ^4(4R -25M) + 8Q^6}{R^5(MR-Q^2)^2}\, u^t }\\
 & & \\
 & & \dsp{ +\, \frac{M^2(R^2 - Q^2)-2MRQ^2+2Q^4}{(MR-Q^2)^2}\, u^{\vf\,2} 
 + \frac{2MR-3Q^2}{R^4}\, \eta u^{\vf}, }\\
 & & \\
\nu& = & \dsp{ \frac{R^2(R-M)(R-3M)+2RQ^2(R-M)}{(MR-Q^2)(R^2-2MR+Q^2)}\, m u^{\vf} + 
  \frac{MR-Q^2}{R^2(R^2-2MR+Q^2)}\, m \eta, }\\ 
 & & \\
\sigma & = & \dsp{ \frac{R(R-M)(R^2-3MR+2Q^2)}{(MR-Q^2)(R^2-2MR+Q^2)}\, m u^t 
 -\, \frac{R^2}{MR-Q^2}\,  m \ve + \frac{qQR}{MR-Q^2}, }\\
 & & \\
\chi & = & \dsp{ \frac{(MR^2-Q^2(2R-M))(R^2(R^2-4MR+5M^2) + 2Q^2(R^2-3MR)+2Q^4)}{(MR-Q^2)^2(R^2-2MR+Q^2)^2}\, 
 m u^{\vf} u^t }\\
 & & \\
 & & \dsp{ -\, \frac{MR^2(3R-4M)-Q^2R(4R-7M)-2Q^4}{R^3(R^2-2MR+Q^2)^2}\, m \eta u^t }\\
 & & \\
 & & \dsp{ -\, \frac{R(MR-2Q^2)}{(MR-Q^2)^2} m \ve u^{\vf} - \frac{qQ^3 u^{\vf} }{(MR-Q^2)^2}. }
\ea
\]
Of course, the corresponding expressions for Schwarzschild space-time are obtained automatically by taking $Q = 0$.

\np
\section{Circular orbits with non-minimal hamiltonian \label{ab}} 

In section \ref{ps7} we derived an equation for circular orbits for test particles in the presence of
Stern-Gerlach interactions, eq.\ (\ref{7.8}). In this appendix we give the explicit expressions for the 
quantities $A$, $B$ and $C$, which are polynomials in the angular velocity velocity $u^{\vf}$ for 
given values of radius $R$ and angular momentum per unit of mass $\eta$. They read as follows:
\be
\ba{lll}
A & = & \dsp{ \frac{M (\eta - R^2 u^{\vf})}{R^2} \left\{ \frac{6M^2 \eta}{R^3} \lh 1 - \frac{2M}{R} \rh
 \lh 2 - \frac{m}{M} \rh  \rd }\\
 & & \\
 & & \dsp{ -\, 4M u^{\vf} \left[ 3 \lh 1 - \frac{2M}{R} \rh \lh 1 - \frac{M}{R} - \frac{m}{2R} \rh + 
   \frac{M\eta^2}{R^3} \lh 1 - \frac{3M}{R} \rh \right] }\\
 & & \\
 & & \dsp{ +\, \eta R u^{\vf\,2} \left[ \frac{2M}{R} \lh 7 - \frac{16M}{R} \rh - \frac{3m}{M} 
   \lh 1 - \frac{2M}{R} \rh \lh 1 - \frac{4M}{R} + \frac{6M^2}{R^2} \rh \right] }\\
  & & \\
  & & \dsp{ \ld + 12 R^3 u^{\vf\,3} \left[ 1 - \frac{49M}{6R} + \frac{19M^2}{R^2} - \frac{13M^3}{R^3}
   + \frac{m}{4M} \lh 1 - \frac{6M}{R} + \frac{14M^2}{R^2} - \frac{12M^3}{R^3} \rh \right] \right\}, }\\
 & & \\
B & = & \dsp{ \frac{12M(\eta - R^2 u^{\vf})^2}{R^3} \left\{ \frac{3M^2}{R^2} \lh 1 - \frac{2M}{R} \rh 
 + \frac{2M^3 \eta^2}{R^5} \rd }\\
 & & \\
 & & \dsp{ - \frac{M\eta}{R}\, u^{\vf} \left[ \frac{M}{R} \lh 5 - \frac{6M}{R} \rh 
  - \frac{3m}{M} \lh 1 - \frac{2M}{R} \rh^2 \right] }\\
 & & \\
 & & \dsp{ \ld - MR u^{\vf\,2} \left[ 3 - \frac{20M}{R} + \frac{26M^2}{R^2} 
  + \frac{3m}{M} \lh 1 - \frac{2M}{R} \rh^2 \right] \right\}, }\\
 & & \\ 
C & = & \dsp{ \frac{72 M^3 (\eta - R^2 u^{\vf})^4}{R^7} \left[ \frac{M}{R} 
 - \frac{3m}{2M} \lh 1 - \frac{2M}{R} \rh \right]. }
\ea
\label{a2.1}
\ee

\np

\bibliographystyle{newutphys}
\bibliography{bibliography_spinletter}

\end{document}